\documentclass[12pt,twocolumn,english]{emulateapj}
\usepackage{natbib,graphicx,epstopdf}

\usepackage[T1]{fontenc}
\usepackage[latin9]{inputenc}
\usepackage{float}
\usepackage{graphicx}

\makeatletter

\providecommand{\tabularnewline}{\\}

\makeatother

\usepackage{babel}

\begin{document}

\title{SECONDARY ECLIPSE PHOTOMETRY OF THE EXOPLANET WASP-5b WITH WARM \textit{SPITZER}}

\author{Nathaniel J. Baskin\altaffilmark{1}}
\author{Heather A. Knutson\altaffilmark{1}}
\author{Adam Burrows\altaffilmark{2}}
\author{Jonathan J. Fortney\altaffilmark{3}}
\author{Nikole K. Lewis\altaffilmark{4,11}}
\author{Eric Agol\altaffilmark{5}}
\author{David Charbonneau\altaffilmark{7}}
\author{Nicolas B. Cowan\altaffilmark{6}}
\author{Drake Deming\altaffilmark{8}}
\author{Jean-Michel Desert\altaffilmark{1}}
\author{Jonathan Langton\altaffilmark{9}}
\author{Gregory Laughlin\altaffilmark{3}}
\author{Adam P. Showman\altaffilmark{10}}

\affil{\altaffilmark{1}Division of Geological and Planetary Sciences, California Institute of Technology, Pasadena, CA 91125, USA}
\affil{\altaffilmark{2}Department of Astrophysical Sciences, Princeton University, Princeton, NJ 05844, USA}
\affil{\altaffilmark{3}Department of Astronomy and Astrophysics, University of California at Santa Cruz, Santa Cruz, CA 95064, USA}
\affil{\altaffilmark{4}Department of Earth, Atmospheric and Planetary Sciences, Massachusetts Institute of Technology, Cambridge, MA 02139, USA}
\affil{\altaffilmark{5}Department of Astronomy, University of Washington, Box 351580, Seattle, WA 98195, USA}
\affil{\altaffilmark{6}Center for Interdisciplinary Exploration and Research in Astrophysics, Northwestern University, Evanston, IL 60208, USA}
\affil{\altaffilmark{7}Harvard-Smithsonian Center for Astrophysics, Cambridge, MA 02138, USA}
\affil{\altaffilmark{8}Department of Astronomy, University of Maryland, College Park, MD 20742, USA}
\affil{\altaffilmark{9}Department of Physics, Principia College, Elsah IL 62028, USA}
\affil{\altaffilmark{10}Lunar and Planetary Laboratory, University of Arizona, Tucson, AZ 85721, USA}
\affil{\altaffilmark{11}Sagan Fellow}

\date{}

\begin{abstract}

We present secondary eclipse photometry of the extrasolar planet WASP-5b taken in the $3.6$ and $4.5\, \micron$ bands with the \textit{Spitzer Space Telescope}'s Infrared Array Camera as part of the extended warm mission. By estimating the depth of the secondary eclipse in these two bands we can place constraints on the planet's atmospheric pressure-temperature profile and chemistry.  We measure secondary eclipse depths of $0.197\% \pm 0.028\%$ and $0.237\% \pm 0.024\%$ in the $3.6\,\micron$ and $4.5\,\micron $ bands, respectively. For the case of a solar-composition atmosphere and chemistry in local thermal equilibrium, our observations are best matched by models showing a hot dayside and, depending on our choice of model, a weak thermal inversion or no inversion at all.  We measure a mean offset from the predicted center of eclipse of $3.7 \pm 1.8$ minutes, corresponding to $e\cos\omega = 0.0025 \pm 0.0012$ and consistent with a circular orbit.  We conclude that the planet's orbit is unlikely to have been perturbed by interactions with another body in the system as claimed by \citet{fukui11}. 
\end{abstract}


\maketitle

\textwidth=7.1in
\columnsep=0.3125in
\parindent=0.125in
\voffset=-20mm
\hoffset=-7.5mm

\topmargin=0in
\headheight=.15in
\headsep=0.5in
\oddsidemargin=0in
\evensidemargin=0in
\parskip=0cm


\textheight=64\baselineskip
\textheight=\baselinestretch\textheight
\ifdim\textheight>25.2cm\textheight=25.0cm\fi

\topskip\baselineskip
\maxdepth\baselineskip

\let\tighten=\relax
\let\tightenlines=\tighten
\let\singlespace=\relax
\let\doublespace=\relax

\def\eqsecnum{
    \@newctr{equation}[section]
    \def\theequation{\hbox{\normalsize\arabic{section}-\arabic{equation}}}}

\def\lefthead#1{\gdef\@versohead{#1}} \lefthead{\relax}
\def\righthead#1{\gdef\@rectohead{#1}} \righthead{\relax}
\let\shorttitle=\lefthead        
\let\shortauthors\righthead      

\section{Introduction}

Hot Jupiters are a class of extrasolar planet that, as the name suggests, are similar in size and composition to
Jupiter but orbit very close to their parent star and have correspondingly high effective temperatures, ranging from 1000K to over 3000K. By measuring the wavelength-dependent decrease in light during the secondary eclipse (when the planet passes behind its star), we can characterize the planet's emission spectrum and deduce its atmospheric properties \citep{deming05, charbonneau05}. These atmospheric properties include: whether or not the atmosphere has a temperature inversion and how well heat is redistributed from the planet's dayside to its nightside. In addition, the relative timing of transits and secondary eclipses can constrain the eccentricity of the planet's orbit. 

To date \textit{Spitzer} has measured secondary eclipses for nearly fifty extrasolar planets.  The resulting studies indicate that hot Jupiters can be differentiated by the presence or absence of a strong thermal inversion in the planet's upper atmosphere \citep{burrows08, fortney08, barman08, mahusudhan09}. Although it exhausted the last of its cryogen in 2009, the \textit{Spitzer Space Telescope}'s Infrared Array Camera (IRAC) remains functional in its $3.6$ and $4.5\;\micron $ bands \citep{fazio04}, and the telescope has continued to survey the emission spectra of hot Jupiters as part of \textit{Spitzer}'s extended warm mission.

Some groups have suggested that absorbers such as TiO in the upper atmosphere of hot Jupiters are responsible for atmospheric inversions \citep{hubeny03, fortney08}.  However, large-scale atmospheric mixing would be required to preserve gaseous TiO in the upper atmosphere, as this molecule should condense in the deep interiors of most hot Jupiters.  It is uncertain whether such macroscopic mixing should take place in a stably stratified atmosphere \citep{showman09, spiegel09, parmentier13}.  In addition, temperature inversions have been observed on planets like XO-1b \citep{machalek08} that have dayside temperatures below the condensation point of TiO.  Sulfur-containing compounds have been proposed as an alternative absorber \citep{zahnle09}.

In this paper, we present observations of the transiting hot Jupiter WASP-5b. This planet is very dense compared to other planets in its class, suggesting the presence of a large, metal-rich core \citep{anderson08, southworth09, mahusudhan09, gillon09, dragomir11, fukui11, pont11, hoyer12}.  Knutson et al. 2010 reported evidence for a correlation between stellar activity and hot Jupiter emission spectra, where hot Jupiters with strong temperature inversions tend to orbit more quiet stars and more active stars typically host planets without inversions.  This may be due to the fact that the more intense UV radiation from active stars destroys the compounds that are responsible for creating thermal inversions.  WASP-5 is a modestly active G4V star with $log(R') = -4.75$ and $S_{HK}=0.215$, so it is not expected that this planet would have a strong temperature inversion.  In addition, papers such as \cite{cowan11} , \cite{showman02}, and \cite{perna12} find that planets hotter than approximately 2000~K have weak transport of energy to the nightside. With a predicted equilibrium temperature of $1720$ K for the case of zero albedo and full day-night redistribution of energy, WASP-5b is near the boundary of this transition.  In this paper, we will test the correlations proposed by \cite{knutson10} and \cite{cowan11} by constraining the atmospheric properties of WASP-5b.

In a previous study, \cite{fukui11} found that the intervals between WASP-5b's transits do not appear to be constant, which could indicate the presence of an additional body in the system perturbing the planet's orbit.    Hoyer et al. (2012), however, disputes the claims of such transit timing variations (TTVs). In Section~\ref{Orbital Eccentricity}, we present the first measurements of this planet's secondary eclipse times and discuss the implications of our measurements for the proposed perturber.

\begin{figure}[t]

\noindent \begin{centering}
\epsscale{1.15}
\plotone{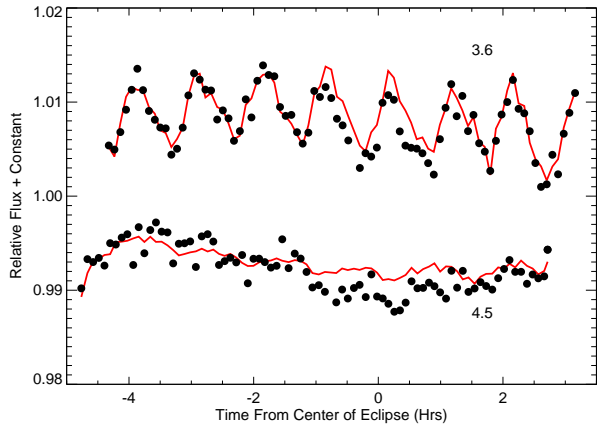}
\par\end{centering}

\caption{Raw photometry at $3.6$ and $4.5\;\micron $ vs time from the center of the predicted secondary eclipse.  The decorrelation functions to correct for intrapixel sensitivity are overplotted in red.  Data are binned in sets of 25 points.}
\label{fig1}
\end{figure}

\section{Observations and Methods}

We began by extracting photometry from calibrated images in the $3.6\, \micron$ and $4.5\, \micron$ bands from the Spitzer Space Telescope following the methods described in \citet{knutson12} and \citet{lewis13}.  Each channel consists of $2115$ 12~s images acquired over a period of $7.7$ hr.  The basic calibrated data (BCD) files used were dark-subtracted, linearized, flat-fielded, and flux-calibrated. We correct for transient ``hot pixels'' in a $20\times20$ pixel box around the star by removing intensity values $>3\sigma$ from the median value at that pixel position in the surrounding 50 images. These values are then replaced by the median.  In addition, to further reduce the noise in the data we trimmed the first 53 minutes from channel 1 ($3.6\, \micron$) data and the first 15 minutes from channel 2 ($3.6\, \micron$) data, which exhibit larger deviations in position.   As a test we repeated our fits with no trimming and found that our best-fit eclipse depths change by less than $1\sigma$.

We performed aperture photometry using the aper routine from the IDL Astronomy Library (http://idlastro.gsfc.nasa.gov/homepage.html) with radii ranging from $2.0$ to $3.0$ pixels in intervals of $0.1$ pixels, and from $3.0$ to $5.0$ pixels in half-pixel intervals.  We found that setting the aperture for channel 2 equal to $2.2$ pixels produced the smallest root mean square (rms) scatter in the data. We obtained superior results in channel 1 using a time-varying aperture proportional to the square root of the noise pixel value \citep{knutson12, lewis12} minus $0.1$ pixels.  In channel 2, however, using a time varying aperture increased the scatter, so a fixed size aperture was selected. Our median aperture size for channel 1 was 1.7 pixels.  We determined the position of the star in both channels using flux-weighted centroiding with a radius of 5.0 pixels.

\begin{table*}[t]
\centering
\caption{Best-fit values for the secondary eclipse depth and time$^1$}
\begin{tabular}{cccccccccccc}
\hline 
\hline
Wavelength ($\micron $) &  & Eclipse Depth (\%) &  & Brightness Temperature (K) &  & Center of Eclipse (BJD) &  & Eclipse Offset (hours)\tabularnewline
\hline 
$3.6$ &  & $0.197\pm0.028$ & & $2210\pm140$ & & $2455200.4279\pm0.0016$ & & $4.0\pm2.4$\tabularnewline
$4.5$ &  & $0.237\pm0.024$ &  & $2130\pm110$ & & $2455174.3726\pm0.0019$ &  & $3.3\pm2.8$\tabularnewline
\hline 
\end{tabular}
\label{tab1}
\end{table*}

\begin{figure}[t]
$\;$

\noindent \begin{centering}
\epsscale{1.15}
\plotone{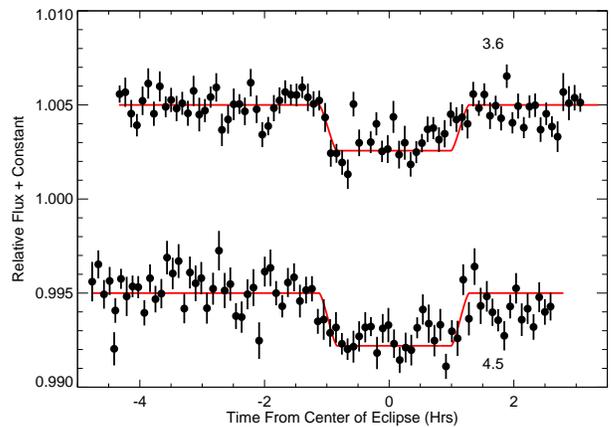}
\par\end{centering}

\caption{Photometry and best-fit eclipse curve for both wavebands after decorrelation versus time from the predicted center of the secondary eclipse. The data are binned in sets of 25 points, with errors calculated as the standard deviation of the points in each bin divided by the square root of the number of points.}
\label{fig2}
\end{figure}

We next correct for the dominant instrument effect in the two bands. Since the sensitivity of individual pixels varies from the center out to the edge, the apparent flux of the star will fluctuate with the movement of the telescope. To correct for this intrapixel sensitivity variation, we fit the data with quadratic functions in $x$ and $y$ position.  For channel 1, the measured flux $f$ is given by
\noindent \begin{center}\small
$f=f_{0} (c_{1}(x-x_{0})^{2}+c_{2}(y-y_{0})^{2}+c_{3}(x-x_{0})+c_{4}(y-y_{0})+c_{5})$
\end{center}
where $f_0$ is the incident flux, $x$ and $y$ are the positions of the star on the array, $x_{0}$ and $y_{0}$ are
the median positions over the time series, and $c_{1-5}$ are free parameters.   We also tried fitting cubic and linear functions, as well as adding an $xy$ term, but found that a quadratic function yielded the lowest value of the Bayesian Information Criterion (BIC) in both channels.  Figure~\ref{fig1} shows the raw data together with our best-fit quadratic functions.  Figure~\ref{fig2} displays the normalized data with best-fit eclipse models for each channel.

We determined our best-fit parameters from a $\chi^2$ minimization and used a Markov Chain Monte Carlo (MCMC) method \citep{ford05, winn07} with $10^{5}$ steps to determine the corresponding uncertainties in our fitted parameters.  We set the uncertainty on individual points in our time series equal to the standard deviation of the best-fit residuals in each band (0.37\% and 0.43\% in channels 1 and 2, respectively).   We find that the RMS of our best-fit residuals is a factor of 1.2 (channel 1) and 1.1 (channel 2) times higher than the photon noise limit.  We used seven free parameters in each channel, including the eclipse depths, timing offsets, and the intrapixel sensitivity corrections. The planetary and stellar radii, orbital period, and orbital inclination were set to the values given in \citet{hoyer12}.  Once the chain is completed we find the point where the $\chi^2$ value for the chain first moves below the median $\chi^2$ value over the entire chain, and trim the steps up to that point.  We define the uncertainty for each parameter as the symmetric range aover which the probability distribution contains $68\%$ of the points around the median.  The distributions for all parameters were approximately Gaussian, and the eclipse depths and times were not strongly correlated with any of the other fit parameters.  Our probability distributions for the eclipse depths and times were smooth and well-sampled, indicating that the chains had fully converged.  See Table~\ref{tab1} for the best-fit eclipse depths and times.\footnote{Both eclipse times are given in the BJD$_{UTC}$ time standard; to convert to BJD$_{TT}$ simply add 66.184 s.  We calculate the timing offsets using the ephemeris from \citet{fukui11} and accounting for the 26.78 s light-travel time in this system \citep{loeb05}.  We also include the uncertainty in the predicted eclipse times at the epoch of our observations, but this is only 39 s and has a negligible effect on the combined uncertainty.}

We also calculate error bars using the ``prayer-bead'' method \citet{gillon09}, which involves shifting the residuals of the data in single point increments and recalculating the best-fit eclipse depth and time.  For a comparison of this method versus other approaches for treating time-correlated noise, see \citet{carter09}.  We find that the prayer-bead method yields error bars consistent with those of the MCMC, and we chose to use the larger of the two errors in each case. For channel 1, we used the prayer-bead error for both the eclipse depth and center of eclipse time (0.028\% and 0.039 hours, respectively) as opposed to the MCMC error (0.027\% and 0.038 hours). For channel 2, we used the MCMC for the eclipse depth (0.024\% as opposed to 0.022\% from the prayer-bead) and both methods give errors of 0.046 hours for the eclipse time.  

\section{Discussion}
\subsection{Atmospheric Temperature Structure}\label{Atmospheric Temperature Structure}
\begin{figure}[t]
$\;$

\noindent \begin{centering}
\epsscale{1.15}
\plotone{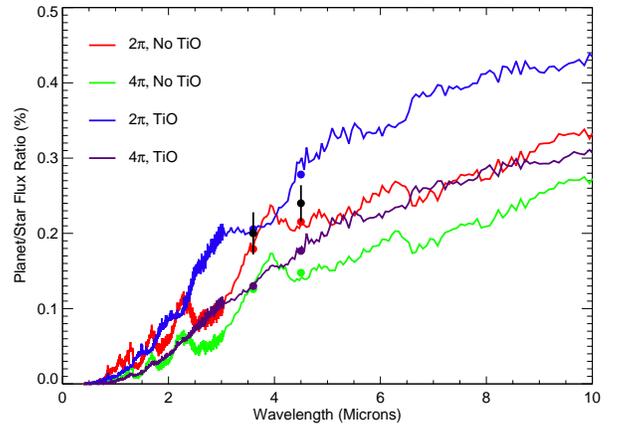}
\par\end{centering}

\caption{Dayside planet/star flux ratio vs wavelength for four model atmospheres (Fortney et al. 2008) with the band-averaged flux ratios for each model superposed (colored circles).  Stellar fluxes were calculated using a 5700~K \texttt{PHOENIX} stellar atmosphere model \citep{hauschildt99}.  The observed contrast ratios are overplotted as the black circles, with uncertainties shown. Two models contain gas-phase TIO, which produces a high-altitude temperature inversion \citep{parmentier13}. In the figure, the red and blue models represent a planet that radiates heat only from the dayside.  The red and blue models represent a planet that radiates heat only from the dayside, while the green and purple models distribute heat equally between the two hemispheres.}
\label{fig3}
\end{figure}

We consider two classes of atmospheric models in this paper.  The first, following the methods of \citet{fortney08} examines the impact of TiO as an absorber in the upper atmosphere.  These models parameterize the redistribution of heat by adjusting the incident flux at the top of the atmosphere.  In Figure~\ref{fig3}, four models are shown.  The red and blue models represent a planet with a hot dayside that redistributes little heat to the nightside.  The green and purple models characterize an atmosphere in which heat is evenly distributed between the planet's day and night sides.  The blue and purple models contain gas-phase TiO, which acts as an absorber in the planet's upper atmosphere.  The red and green models contain no TiO.  We find that the slope between the band averaged flux ratios on the blue model is too steep to fit both of our observed values.  As seen in the figure, our observations are best matched by the red model, which has no thermal inversion and poor redistribution of energy from the dayside to the nightside.  

\begin{figure}[t]
$\;$

\noindent \begin{centering}
\epsscale{1.15}
\plotone{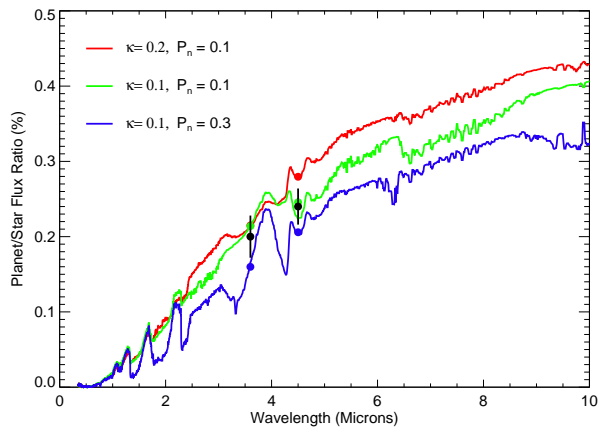}
\par\end{centering}

\caption{Dayside planet/star flux ratio vs wavelength for three model atmospheres (Burrows et al. 2008) with the band-averaged flux ratios for each model superposed (colored circles).  Stellar fluxes were calculated using a 5700~K \texttt{ATLAS} stellar atmosphere model \citep{kurucz05}.  The observed contrast ratios are overplotted as the black circles, with uncertainties shown.  The model parameter $\kappa$ is related to the atmosphere's opacity, while $p$ is related to the heat redistribution between the day and night sides of the planet ($P_n = 0.0$ indicates no heat redistribution, and $P_n=0.5$ indicates complete redistribution)}
\label{fig4}
\end{figure}

The second class of models follow the methods of \cite{burrows08}.  As opposed to modifying the atmospheric levels of TiO, this method utilizes a nonspecific gray absorber at low pressures.  A heat sink is added at depth to allow for the redistribution of heat from the planet's dayside to its nightside.  We use the dimensionless parameter $P_n$ to specify the heat redistribution, where $P_n = 0.5$ indicates evenly distributed energy among both the day and night sides of the planet, and $P_n = 0.0$ indicates no heat redistribution.  This model also puts an extra absorber with an optical opacity, $\kappa$, in the upper atmosphere to heat it by partial absorption of the incident stellar fluxin visible wavelengths. An enhancement in optical wavelength opacity in the planet's atmosphere will produce a thermal inversion, but importantly raises the temperature of the outer atmosphere to better reproduce the photometric signatures of a subset of measured hot Jupiters. The character of such an absorber, and whether it is a molecule or haze, are currently unknown..  Figure~\ref{fig4} shows three of these models.  Our observations are best matched by the green model, which indicates a weak temperature inversion and little heat distribution between the sides of the planet, in good agreement with the \cite{fortney08} models.

\begin{figure}[t]
$\;$

\noindent
\epsscale{1.1}
\plotone{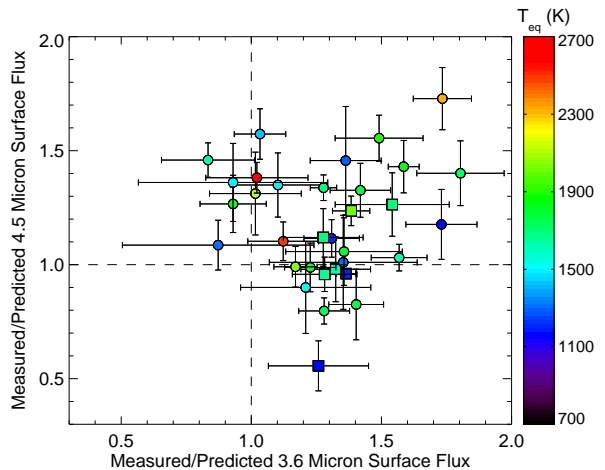}

\caption{Ratio of measured to predicted planet surface flux in the 3.6 vs. 4.5~\micron  \emph{Spitzer} bands for all planets with published secondary eclipse detections at these two wavelengths.  The colors of the points reflect the predicted equilibrium planet temperature $T_{eq}$, which is calculated assuming that the planet has an albedo of zero and re-radiates energy uniformly over its entire surface.  Predicted fluxes are calculated assuming the planet radiates as a blackbody at this temperature.  Measured fluxes are calculated using published secondary eclipse depths and assuming a PHOENIX atmosphere model for the star.  For planets with eccentricities greater than 0.05 measured at 3$\sigma$ significance or higher we show the orbit-averaged temperature.  Planets orbiting active stars (defined as having log(\ensuremath{R'_{\mbox{\scriptsize HK}}}) \citep{knutson10} greater than $-4.9$ and a stellar effective temperature less than 6200 K) are shown as filled squares, all other planets are plotted as filled circles.}
\label{fig5}
\end{figure}

We calculate brightness temperatures for the planet in both bands using a PHOENIX model atmosphere for the star with stellar parameters from \citet{doyle12} and find planetary brightness temperatures of $2210\pm140$~K at 3.6~\micron~and $2090\pm120$~K at 4.5~\micron.  If the planet has an albedo of zero and re-radiates the absorbed flux uniformly over its entire surface, we calculate a predicted equilibrium temperature of 1700~K.  If it re-radiates this energy from the dayside hemisphere alone, this temperature increases to 2030~K.  The measured brightness temperatures therefore provide a model-independent indication that WASP-5b has a hot dayside and relatively inefficient day-night recirculation.  We consider whether or not this is typical of the class of hot Jupiters by plotting the measured surface fluxes for all planets with published secondary eclipses in the 3.6 and 4.5~\micron~\emph{Spitzer} bands and comparing these fluxes to the values for a blackbody at the predicted equilibrium temperature (see Figure~\ref{fig5}).  We find that these planets universally exhibit flux excesses relative to the coolest predicted blackbody (corresponding to efficient day-night circulation) in at least one, and sometimes both, \emph{Spitzer} bands.  We do not find any clear correlation between flux excesses and the predicted planet temperature, although planets orbiting active stars tend to lie towards the lower right region of the plot.

Our results are consistent with the correlation between stellar activity and hot Jupiter emission spectra, as described in \cite{knutson10}.  According to this study, hot Jupiters with strong temperature inversions tend to orbit more quiet stars, while more active stars typically host planets without inversions.  WASP-5 is a G4V star with $log(R') = -4.75$ and $S_{HK}=0.215$, indicating that the star is modestly active.  This is consistent with our results from the atmospheric models (Section~\ref{Atmospheric Temperature Structure}), which suggest the presence of either a weak thermal inversion or none at all.  In addition, the low heat redistribution of WASP-5b is consistent with the predictions of \cite{cowan11}.

\subsection{Orbital Eccentricity}\label{Orbital Eccentricity}
The timing of the secondary eclipse depends on the eccentricity of the planet's orbit.  As a result, we can utilize the predicted and observed times of eclipse to constrain $e \cos\omega$, where $e$ is the orbital eccentricity and $\omega$ is the argument of pericenter \citep{charbonneau05}.  We note that if the planet has a non-uniform dayside brightness distribution we may measure an apparent offset in the secondary eclipse time; following the model of \cite{cowan11b}, as implemented in \cite{agol10}, we calculate that the maximum expected time offset due to an offset hot spot would be 37 s or less than $1\sigma$~for our measurement.  

We calculate our predicted eclipse times using the ephemeris from \citet{fukui11}, which has smaller errors than the ephemeris reported by \citet{hoyer12}; this may be due to Fukui et al.'s inclusion of both the radial velocity and transit data in their fits.  Using these values and accounting for the 27 s light travel time we find a mean offset from the predicted center of eclipse of $3.7 \pm 1.8$ minutes, translating to an $e\cos\omega$ value of $0.0025 \pm 0.0012$.   This indicates that, unless our line of sight is aligned with the planet's semi-major axis, the orbit is circular.  Our result is also consistent with the 95\% confidence upper limit of 0.026 on the planet's eccentricity from \citet{husnoo12}.  We note that \citet{fukui11} claimed to detect transit timing variations for this planet, although this claim has been disputed by \citet{hoyer12}.  If the planet is being perturbed by an additional body in the system we would expect these perturbations to excite its orbital eccentricity and to maintain that eccentricity despite ongoing circularization from tidal interactions with the parent star.  The lack of any orbital eccentricity therefore provides further evidence against the presence of a perturbing body in this system. 

\section{Conclusions}

We observed secondary eclipses of the extrasolar planet WASP-5b in the $3.6$ and $4.5\,\micron $ bands, and compared our measurements with atmospheric models generated according to the methodology of Burrows et al. 2008 and Fortney et al. 2008.  Our eclipse values of  $0.197\% \pm 0.028\%$ and $0.237\% \pm 0.024\%$ in these bands are best matched by models indicating a weak thermal inversion or no inversion at all.  Our measurements are consistent with the observed correlation between stellar activity and the presence of thermal inversions presented in \cite{knutson10}.  In addition, these one-dimensional atmospheric models all indicate minimal day-night redistribution of energy for WASP-5b, with the Burrows models suggesting a redistribution factor of $0.1$, consistent with the prediction of \cite{cowan11}.  Measurements at additional wavelengths would be helpful to confirm or reject the presence of the inversion.  These results further demonstrate the utility of warm \textit{Spitzer} in characterizing the properties of hot Jupiter atmospheres, despite being limited to the $3.6$ and $4.5\,\micron $ channels.  Additional measurements in the J, H, and K bands would also allow us to constrain the metallicity of WASP-5b's atmosphere, which is likely to be enhanced relative to other hot Jupiters.

By measuring the mean timing offset from the predicted center of eclipse ($3.7 \pm 1.8$ minutes), we calculate the parameter $e\cos\omega$ to be $0.0025 \pm 0.0012$.  This indicates that the planet's orbit must be circular unless our line of sight is well-aligned with the semi-major axis.  These observations, along with those in Hoyer et al. 2012, argue against the possibility of ongoing perturbations from an additional body in the WASP-5 system.

\end{document}